\documentclass[11pt,preprintnumbers,amsmath,amssymb,onecolumn]{revtex4}

\usepackage[version=4]{mhchem}
\usepackage{amsmath}
\usepackage[utf8]{inputenc}
\usepackage[driverfallback=dvipdfm]{hyperref}
\usepackage{color}
\usepackage{siunitx}
\usepackage[inline]{enumitem}
\usepackage{graphicx}
\usepackage{caption}
\usepackage{subcaption}
\graphicspath{ {./home/cbb/tex files} }
\usepackage{makeidx}
\usepackage{bm}
\usepackage[title]{appendix} 
\usepackage{tabu}
\usepackage{amssymb}
\usepackage{float}
\usepackage{lineno}

\begin{document}

\title{\large {Emergent dynamics in an astrocyte-neuronal network coupled \textit{via} nitric oxide } }
\author{Bhanu Sharma$^{1}$, Spandan Kumar$^{2}$, Subhendu Ghosh$^{1}$, Vikram Singh$^{3*}$}
\email{vikramsingh@cuhimachal.ac.in}
\affiliation{$^1$ Department of Biophysics, University of Delhi South Campus, New Delhi, India .\\
$^2$ School of Social Sciences, Indira Gandhi National Open University, New Delhi, India.\\ $^{3}$ Centre for Computational Biology and Bioinformatics, Central University of Himachal Pradesh, Dharamshala, India. }

\begin{abstract}{\noindent}\textbf{Abstract}
In the brain, both neurons and glial cells work in conjunction with each other during information processing. Stimulation of neurons can cause calcium oscillations in astrocytes which in turn can affect neuronal calcium dynamics. The ``glissandi" effect is one such phenomenon, associated with a decrease in infraslow fluctuations, in which synchronized calcium oscillations propagate as a wave in hundreds of astrocytes. Nitric oxide molecules released from the astrocytes contribute to synaptic functions on the basis of the underlying astrocyte-neuron interaction network. In this study, by defining an astrocyte-neuronal (A-N) unit as an integrated circuit of one neuron and one astrocyte, we developed a minimal model of neuronal stimulus-dependent and nitric oxide-mediated emergence of calcium waves in astrocytes. Incorporating inter-unit communication $via$ nitric oxide molecules, a coupled network of 1,000 such A-N units is developed in which multiple stable regimes were found to emerge in astrocytes. We examined the ranges of neuronal stimulus strength and the coupling strength between A-N units that give rise to such dynamical behaviors.  We also report that there exists a range of coupling strength, wherein units not receiving stimulus also start showing oscillations and become synchronized. Our results support the hypothesis that glissandi-like phenomena exhibiting synchronized calcium oscillations in astrocytes help in efficient synaptic transmission by reducing the energy demand of the process.  
\end{abstract}

\maketitle

\section{Introduction}

Procreation of calcium waves in astrocytes from one astrocytic cell to another is termed as ``glissandi", and is reported to be present in hundreds of astrocytes in the form of synchronized calcium oscillations ~\citep{kuga2011large}. Emergence of this phenomena is known to be dependent upon neuronal activity, and is  associated with reduction in infraslow oscillations ~\citep{kuga2011large}, which was further correlated with a decrease of around 42\% in infraslow potential fluctuations. Astrocytes have also been shown to help in reducing energy requirements of neurons upto 60\% by diluting the individual requirements of synapses ~\citep{weber2015astrocyte}. 

Glutamate uptake by astrocytes is an important regulatory mechanism underlying interactions between neurons and astrocytes ~\citep{rose2018astroglial}. At physiological level, astrocytes maintain extracellular concentration of glutamate by clearning excess neurotransmitters from synaptic cleft ~\citep{mahmoud2019astrocytes}.  Previously, it was thought that astrocytes do not have much role in cognition but growing evidences suggest that astrocytes are indeed involved in many higher order brain functions, such as, synaptic transmission ~\citep{sibille2014astroglial,khakh2015diversity}. They are involved in modulating synaptic strength by coupling synaptic connections into functional assemblies, thereby imposing higher level organization in information processing ~\citep{mederos2018astrocyte,santello2019astrocyte}.

This is probably achieved through sensitive dependence of calcium ocsillations inside astrocytes on glutamate released from neurons ~\citep{zonta2002calcium}. These calcium oscillations ~\citep{kastanenka2020roadmap} are found to be dependent upon type and frequency of stimulus recieved by neurons ~\citep{pasti1997intracellular}, their frequency increases when there is frequent firing of neurons. Astrocytes can modulate neurons by inducing changes in their calcium levels as demonstarated in culture studies  ~\citep{nedergaard1994direct}.  This could be due to direct communication between two of them $via$ gap-junctions ~\citep{seifert2006astrocyte, nedergaard2003new} or because of indirect chemical signaling through gliotransmitter release from astrocytes ~\citep{parpura1994glutamate}. 

One such chemical for astrocyte-neuron communication, that can also act as a good coupling agent ~\citep{iadecola1994nitric}, is nitric oxide (NO). Reasons are as follows: 1) It can increase the glutamate production in neurons by binding with soluble guanylyl cyclase (sGC) enzyme ~\citep{wang2017nitric}, 2) It is an endothelium derived relaxing factor (EDRF) which affects almost all cellular processes by performing diverse functions ~\citep{levine2012characterization, forstermann2012nitric}, 3) It can diffuse from its production site through the cell membrane and do not need any specialized release machinery ~\citep{askalan2006astrocytic}, and 4) It is also known to act as a signalling molecule that can cover large distances between distant synapses and cells in the brain ~\citep{yong2006modeling}, and thereby has key role in synaptic plasticity, synaptic transmission, neuronal development as well as in neurotoxicity ~\citep{schuman1994nitric, jaffrey1995nitric}. Synthesis of NO from nitric oxide synthase (NOS) enzyme is also calcium dependent ~\citep{feinstein1994nitric}. Elevation of calcium levels in astrocytes increases NO production ~\citep{publicover1993amplification} and diffusion of NO molecules outside the glial cells, therefore have regulatory effects on calcium waves in nearby cells ~\citep{li2003calcium, munoz2015control}.

Furthermore, experimental studies have reported that an increase in astrocytic network synchronization occurs before neuronal synchronization, thereby suggesting their causative role in generating slow wave activity ~\citep{szabo2017extensive}. In another study, it was found that neuronal synchronization and correlated pairs in hippocampus decreased on injecting astrocytes with BAPTA, which is a calcium chelator ~\citep{sasaki2014astrocyte}. The same study also demonstrated that average neuron-glia distances leading to neuronal synchronization were larger than astrocyte's territory. Even though the authors suggested this to be an effect of astrocyte's morphology, it could also be a result of long diffusing molecules. Motivated by these studies, in the present work, we developed a minimal model to explore the collective behaviour of A-N units as a result of one such long diffusing molecule, nitric oxide (NO). For this, we constructed a network of astrocyte-neuron (A-N) units, in which the stimulus is given to neurons and its effect is observed in astrocytes in the form of cytosolic $Ca^{2+}$ waves. Multistable dynamical regimes were observed in these astrocytic $Ca^{2+}$ oscillations. Multistability is reported to exist in both natural ~\citep{wunsche2005synchronization, nobukawa2020synchronization} as well as synthetic biological systems ~\citep{ullner2007multistability}  at various scales of biological complexity, ranging from ecological to cellular networks. Some are in neuronal dynamical systems ~\citep{huang2008multistability, ngouonkadi2016bifurcations, ronconi2017multiple} and in single neuron  ~\citep{malashchenko2011six}. The detailed model is presented in the next Section. Results are presented and discussed in detail in Section 3 and Section 4, respectively.

\section{The Model}

In this work, we developed a network of all-to-all connected 1,000 A-N units to study the role of NO in astrocyte neuronal communication. Recent findings suggest that the ratio of glial cells to neurons in brain to be around 1:1 ~\citep{von2016search}. It was the guiding factor for designing the fundamental unit of this network consisting of one neuron and one astrocyte.

\begin{figure}[H]
\begin{subfigure}{0.44\textwidth}
  \centering
  \includegraphics[height=4.0cm, width=7.0cm]{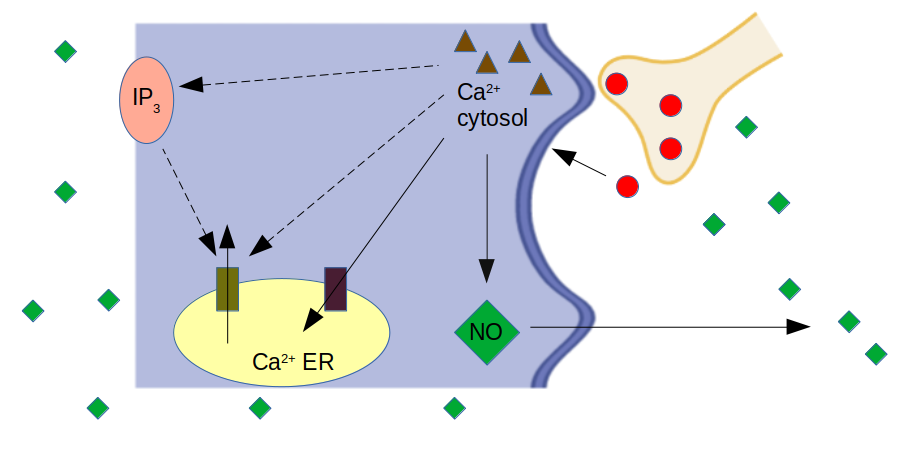}
  \caption{}
  \label{F1a}
  \end{subfigure}%
\begin{subfigure}{0.44\textwidth} 
  \centering
  \includegraphics[height=7.0cm, width=9.3cm]{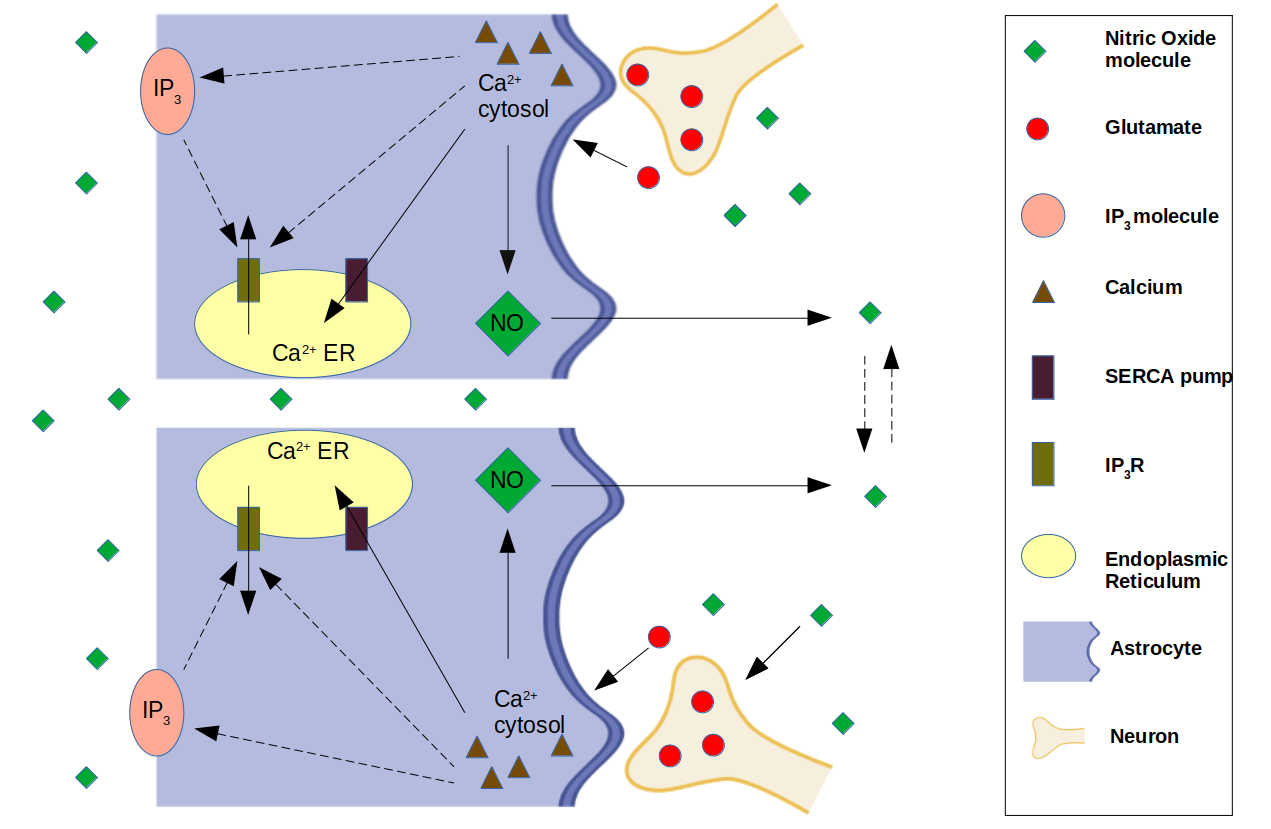}
  \caption{}
  \label{F1b}
  \end{subfigure}
\caption{Schematic representation of the model. (a) Single integrated unit of one astrocyte and one neuron. Neuronal action potential releases glutamate into synaptic cleft, which leads to calcium influx in astrocyte. Calcium enters the ER $via$ SERCA pump, and $IP_{3}$ binds to $IP_{3}R$ allowing calcium flux from ER to cytoplasm. Elevation in calcium concentration further activates the cascade and the production of NO takes place. This NO in turn acts on the neuron and enhances the release of glutamate. (b) Two A-N units are coupled $via$ NO molecules, thereby establishing communication between two units.}
\label{F1}
\end{figure}

For modelling the same, we considered the model of $Ca^{2+}$ oscillations in single astrocyte as given by Lavrentovich \& Hemkin ~\citep{lavrentovich2008mathematical} as the basic framework. This model consists of three variables, viz. (i) calcium concentration in cytosol ($Ca_{cyt}$), (ii) $IP_{3}$ concentration in cytosol ($IP_{3_{cyt}}$), and (iii) calcium concentration in endoplasmic reticulum ($Ca_{ER}$). Biophysics of this model is as follows: Inositol triphosphate ($IP_{3}$) along with calcium ions opens the inositol 1,4,5-trisphosphate receptor ($IP_{3}R$) channel, which allows calcium outflux from endoplasmic reticulum (ER) and increases calcium concentration in the cytoplasm of astrocytes. Calcium ions also positively act on the activation process of phospholipase C (PLC) which in turn enhances the production of $IP_{3}$. Furthermore, if the cytosolic calcium concentration becomes high, it inhibits the $IP_{3}R$ channel. On the other hand, the integration of $Ca^{2+}$ ions in the ER occurs through the Sarco(endo)plasmic reticulum calcium ATPase (SERCA) pump. 

In the single A-N unit as shown in Fig.~\ref{F1a}, proposed in this work, we incorporated the above described astrocytic calcium dynamics model and extended it to incorporate the minimal neuronal circuit, and then created a network of one thousand such units coupled $via$ NO molecule (Fig.~\ref{F1b}).    

In single astrocyte-neuronal unit, each neuron recieves stimulus in the form of a random input ~\citep{softky1993highly} and releases glutamate into the synaptic cleft. Released glutamate triggers the astrocyte activity $via$ 3 routes: 1) by binding with metabotropic receptors, 2) by binding with ionotropic receptors like AMPA, Kianate, and NMDA, and 3) by glutamate transporters ~\citep{verkhratsky2007glutamate}. The third way is used to maintain glutamate homeostasis inside the brain as majority of glutamate released (approximately 80\%) from synaptic neuron is taken up by astrocyte and is recycled ~\citep{swanson2005astrocyte}. In this work, we have used a generic pathway representing first two routes.

Calcium enters into astrocyte through glutamatergic activity as mentioned above, which is modelled as a linear term in the equation (1). This cytosolic $Ca^{2+}$ can bind to $IP_{3}R$ in the presence of $IP_{3}$ and can both increase or inhibit the channel function depending upon calcium concentration in cytosol. Experimental findings suggest that the dependence of $IP_{3}R2$ on $IP_{3}$ is sigmoidal ~\citep{tu2005functional}. 

The calcium ATPase pumps present upon endoplasmic reticulum can force calcium ion from cytosol into ER. Therefore, the change in astrocyte's cytosolic calcium concentration is negatively dependent upon ATPase pumps, modelled using variable B, and positively on calcium induced calcium release mechanism, modelled using variable A, in the equation below:

\begin{equation}
\frac{d(Ca_{cyt,i})}{dt} = k_{g}.G_{i} - k_{out}.Ca_{cyt,i} + A - B + k_{f}.(Ca_{ER,i}-Ca_{cyt,i})
\end{equation}

Here, ($k_{g} .G_{i}$) determines the dependence of cytosolic calcium influx on glutamate ~\citep{eusebi1985post}. The second term ($k_{out}.Ca_{cyt,i}$) signifies the rate of outflux into extracellular environment from the cytosol. Values of rate constants were used from referring ~\citep{hofer2002control, venance1997mechanism}. Calcium leakage in astrocytic cytosol is denoted by $k_{f}.(Ca_{ER,i}-Ca_{cyt,i})$. $k_{CaA}$ and $k_{CaI}$ are the activating and inhibiting constants. 

Dynamics of both variables A and B is sigmoidal in nature and therefore modelled using Hill equation, taking Hill coefficients from experimental studies ~\citep{tu2005functional, tu2005modulation, falcke2004reading}. $\upsilon_{M3}$ is the maximum calcium flux into cytoplasm and $\upsilon_{M2}$ is the maximum calcium flux through SERCA pump. Values of $\upsilon_{M3}$ and $\upsilon_{M2}$ are also taken on the basis of experimental results ~\citep{hofer1999model, falcke2004reading}. 

\begin{equation*}
\begin{aligned}
A&= 4 \upsilon_{M3} \ast ( \frac{k_{CaA}^{n} Ca_{cyt,i}^{n}}{(Ca_{cyt,i}^{n} + k_{CaA}^{n}) (Ca_{cyt,i}^{n} + k_{CaI}^{n})})  \ast  (\frac{{IP_{3_{cyt}}^{m}}}{{IP_{3_{cyt}}^{m}} + k_{ip3}^m}) (Ca_{ER,i} - Ca_{cyt,i}) \\
B&=\upsilon_{M2} (\frac{Ca_{cyt,i}^{2}}{Ca_{cyt,i}^{2} + k_{2}^2}\large)
\end{aligned}
\end{equation*}

Similarly, Calcium dynamics in ER is modelled as:

\begin{equation}
\frac{d(Ca_{ER,i})}{dt}= B - A - k_{f}.(Ca_{ER,i}-Ca_{cyt,i}) 
\end{equation}

The rate of change of $IP_{3}$ is given as: 
  
\begin{equation}
\frac{d(IP_{3_{cyt},i})}{dt} = \frac{v_{p}.Ca_{cyt,i}^{2}}{k_{p}^{2}+Ca_{cyt,i}^{2}} - (k_{deg}.IP_{3_{cyt},i}) 
\end{equation}

Where, the last term in the above equation represents the degradation of $IP_{3}$ in cytosol and the first term explains that $IP_{3}R$ gets activated by change in cytosolic calcium concentration in sigmoidal fashion ~\citep{hofer2002control}. 

The changes in cytosolic $Ca^{2+}$ concentration, as a result of above mentioned preocesses, leads to activation of downstream cascade through which the formation of nitric oxide synthase (NOS) takes place ~\citep{li2003calcium}. NOS produces NO molecules which can freely diffuse across the cell ~\citep{askalan2006astrocytic, zhao1998inducible}. The production of NO iside astrocyte is modelled as follows:

\begin{equation}
\frac{d(N_{i})}{dt}= -k_{3}.N_{i} + (\frac{k_{pp}.Ca_{cyt,i}}{1+Ca_{cyt,i}}) - \eta N_{i}
\end{equation}

In the above equation, $-k_{3}.N_{i}$ defines NO degradation. The second term represents the production of NO from cytosolic calcium modeled using standard Michaelis–Menten kinetics, as propsed by Comerford et al. ~\citep{comerford2008endothelial}  and $\eta N_{i}$ represents the outflux of NO from the astrocyte's cytosol. As a result of this outflow of NO, its extracellular concencentration ($N_{out}$) starts increasing proportionally to the number of A-N units (P). $N_{out}$ binds to soluble guanylyl cyclase ~\citep{roy2008enzyme, bartus2013cellular}, present in boutons of neurons, which in turn can enhance the  glutamate production.

\begin{equation}
\begin{aligned}
  \frac{dN_{out}}{dt}&= -k_{nout}.N_{out} + \eta \sum_{j=1}^P {N_{j}} - \zeta P N_{out}  \\
  \frac{dN_{out}}{dt}&= -k_{nout}.N_{out} + \eta P \bar{N} - \zeta P N_{out} 
\end{aligned}
\end{equation}

where 
\begin{equation}
 \bar{N} = \frac{1}{P} \sum_{j=1}^P {N_{j}} \\
\end{equation}

At quasi steady state,

\begin{equation}
\begin{aligned}
   (k_{nout} + P\zeta) N_{out}&= \eta P \bar{N} \\
   N_{out}&= \eta_{ext} Q \bar{N} 
\end{aligned}
\end{equation}

where

\begin{equation}
\begin{aligned}
   \eta_{ext}&= \frac{\eta}{\zeta} \\	
   Q&= \frac{P}{(\frac{k_{nout}}{\zeta})+ P}
\end{aligned}
\end{equation}

$Q$ is the coupling strength, which is defined as the strength of connection between two A-N units coupled $via$ NO molecule. $N_{out}$ represents the extracellular NO concentration, $k_{nout}$ is the decay constant for extracellular NO.

In the proposed model, stimulus activates synaptic vesicles to release glutamate ~\citep{parpura2000physiological}. Although we have taken a generalised term for stimulus, it can be interpreted as calcium. Calcium (stimulus) ranges from 0.12 to 0.18 $\mu M sec^{-1}$, which is considered to be in physiological range ~\citep{gleichmann2011neuronal}. To keep our simulations within this range, we have taken the value of stimulus to be 0.17 $\mu M sec^{-1}$. As mentioned above, NO also influences the production of glutamate by binding with sGC. The equation explaining dynamics of glutamate is as follows:

\begin{equation}
\frac{d(G_{i})}{dt}= -k_{glut}.G_{i} + \frac{g.N_{out}^{2}}{1+N_{out}^{2}} + stimulus 
\end{equation}

Where, ($k_{glut}.G_{i}$) represents the decaying of glutamate from $i$ neuronal cells. Second term represents NO mediated release of glutamate into the synaptic cleft.  

\vspace{0.8cm}

All calculations were implemented and performed in the python programing language. ODEs as described in the above equations were developed using law of mass action and Michaelis–Menten kinetics ~\citep{singh2015modelling} and solved  using the $solve$-$ivp$ solver. The initial values of the model were obtained from single astrocytic model proposed in  ~\citep{lavrentovich2008mathematical}. To quantify the extent of synchronization, a parameter  $R$ is defined as per ~\citep{garcia2004modeling}, as given below:

\begin{equation}
M(t)= (\frac{1}{P})  \sum_{i=1}^P Ca_{cyt,i}(t)  
\end{equation}

\begin{equation}
R= \frac{<M^{2}> - <M>^{2}}{\overline{<{Ca_{cyt,i}^{2}> - <Ca_{cyt,i}>^{2}}}}   
\end{equation}

\begin{figure}[H]
\centering
\includegraphics[height=9.0cm, width=16.0cm]{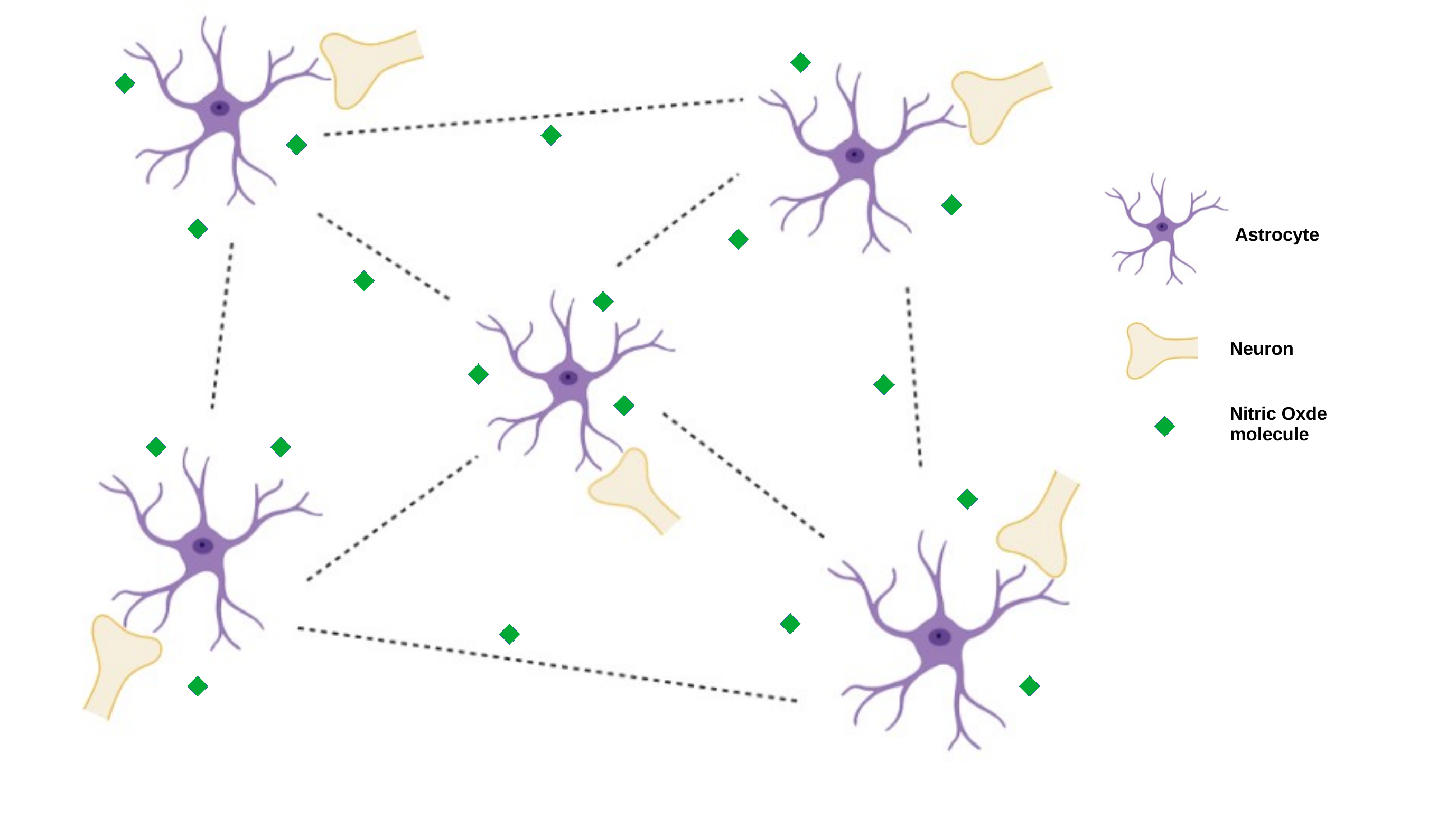}
\caption{An schematic illustration of the network consisting of all-to-all interacting astocyte-neuronal units. Communication is established through nitric oxide molecules. The above figure is prepared using template of the Biorender.}
\label{F2}
\end{figure}

\section{Computational Results}

We simulated all-to-all connected networks of coupled astrocyte-neuronal units under various scenarios of varying stimulus and coupling strength (Q). The range of parameter values used in this model are the same as used in ~\citep{lavrentovich2008mathematical}. Variables used in the model and their intital conditions are shown in Table ~\ref{Table1} as shown below:

\begin{table*}[h]
\begin{center}
\caption{Initial conditions of variables used in the model}
\label{Table1}
\begin{tabular}{ |p{5cm}|p{5cm}|p{5cm}|}
\hline
\textbf{Variable} & \textbf{Initial value (micromolar)}\\
\hline
[$G_{i}$] & 0.1 \\
\hline
[$Ca_{cyt,i}$] & 0.1 \\
\hline
[$Ca_{ER},i$] & 1.0  \\
\hline
[$IP_{3,i}$] & 0.1 \\
\hline
[$N_{i}$] & 0.01 \\
\hline
\end{tabular}
\end{center}
\end{table*}

\subsection{Optimizing stimulus strength in a network of 100 A-N units:}

\begin{figure}[H]
\begin{subfigure}{0.45\textwidth} 
  \centering
  \includegraphics[scale=0.46]{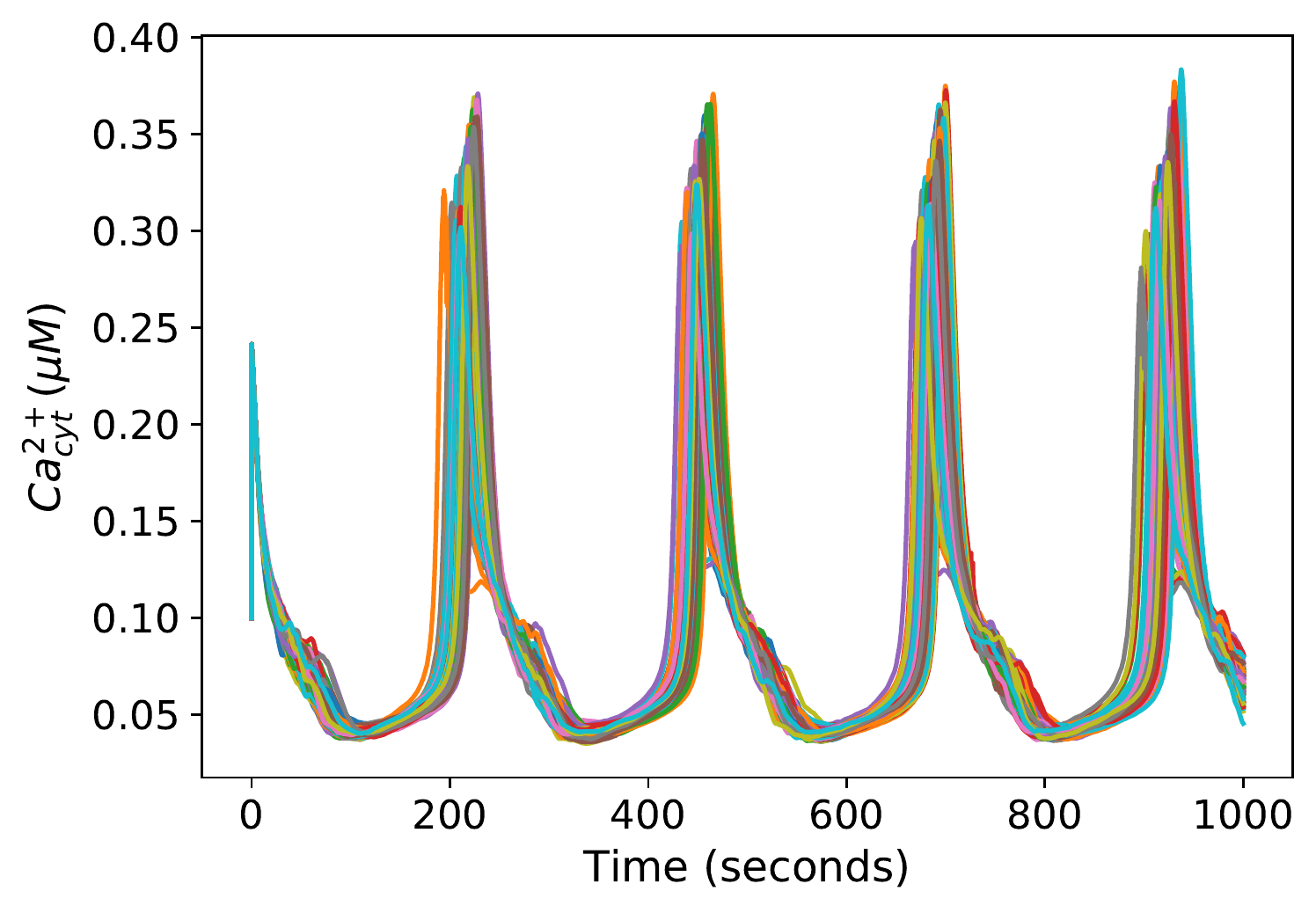}
  \caption{Variation of [$Ca^{2+}$] with time}
  \label{F3a}
  \end{subfigure}%
\begin{subfigure}{0.45\textwidth}
  \centering
  \includegraphics[scale=0.28]{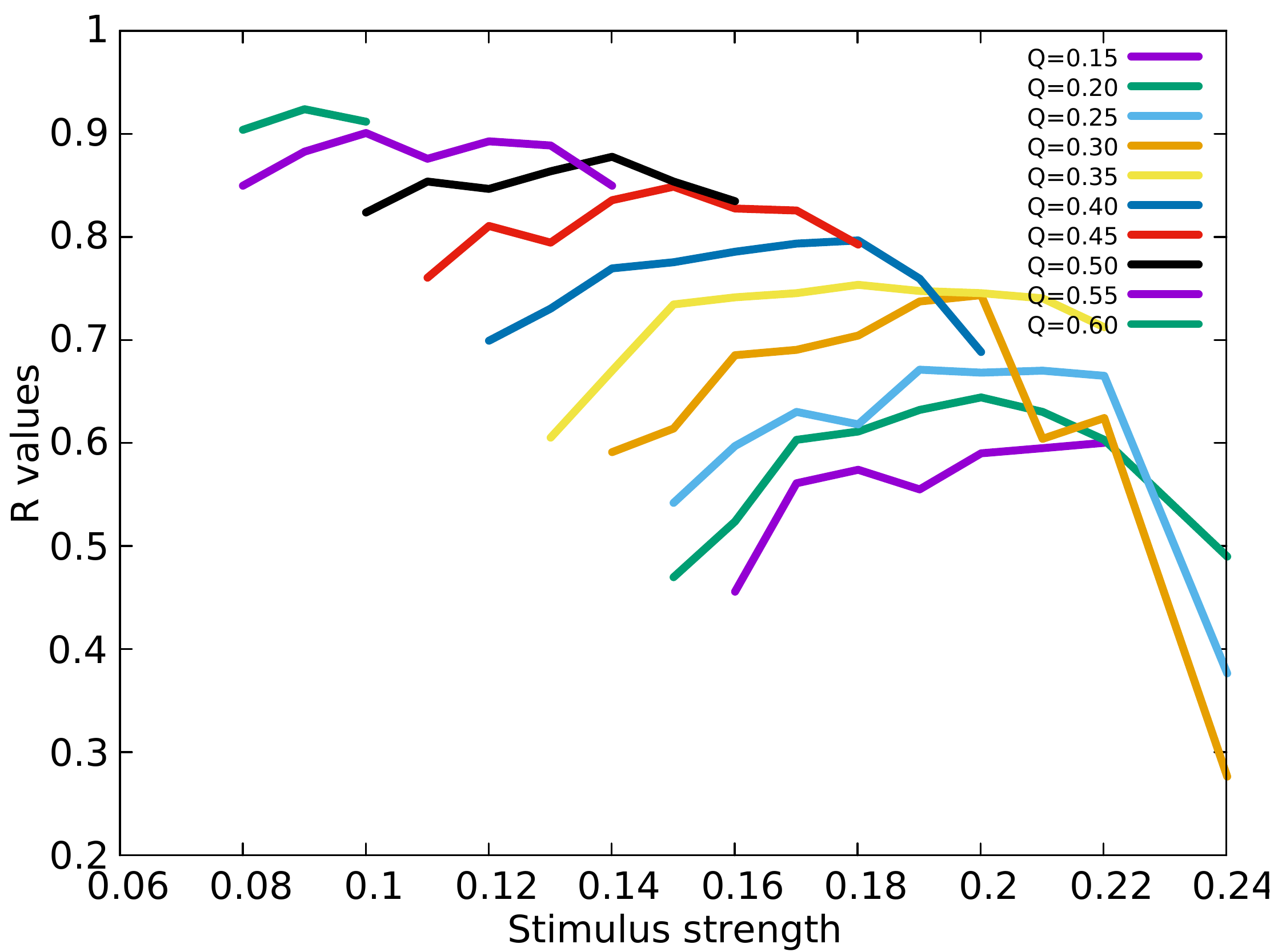}
  \caption{$R$ dependence on $Q$ and stimulus}
  \label{F3b}
  \end{subfigure}
\caption{(a) Variation of astrocytic $Ca^{2+}$ concentration with time. Synchronized $Ca^{2+}$ oscillations in 100 astrocytic cells of 100 A-N units are shown when values of $Q$ and stimulus values are fixed at 0.45 and 0.17 $\mu M sec^{-1}$, respectively. The $Ca^{2+}$ oscillations in each astrocyte of a unit is represented by different colors. (b) Synchronization dependence on stimulus and coupling strength is shown where each line color represents a separate $Q$ value.} 
\label{F3} 
\end{figure}

Stimulus amplitude, given as random pulses were varied from 0.10 to 0.20 $\mu M sec^{-1}$ to check its effect on synchronization of inter-unit astrocytic dynamics $Ca^{2+}$. Fig.~\ref{F3a}  shows one such case in a coupled network of 100 A-N units with coupling strength ($Q$) and stimulus strength fixed at values 0.45 and 0.17 $\mu M sec^{-1}$, respectively. Fig.~\ref{F3b} summarizes the variation observed in synchronization with varying stimulus.  Each curve in Fig.~\ref{F3b} has a shape similar to an inverted V, thus, suggesting an optimal value of stimulus where the cytosolic calcium synchronization is the highest. Oscillations die out beyond the ranges plotted in Fig.~\ref{F3b}, hence, the synchronization parameter ($R$) for those values is not plotted.

\subsection{Comparison of four cases with varying stimulus and coupling strength among A-N units:}

\begin{figure}[H]
\begin{subfigure}{0.44\textwidth}
  \centering
  \includegraphics[scale=0.45]{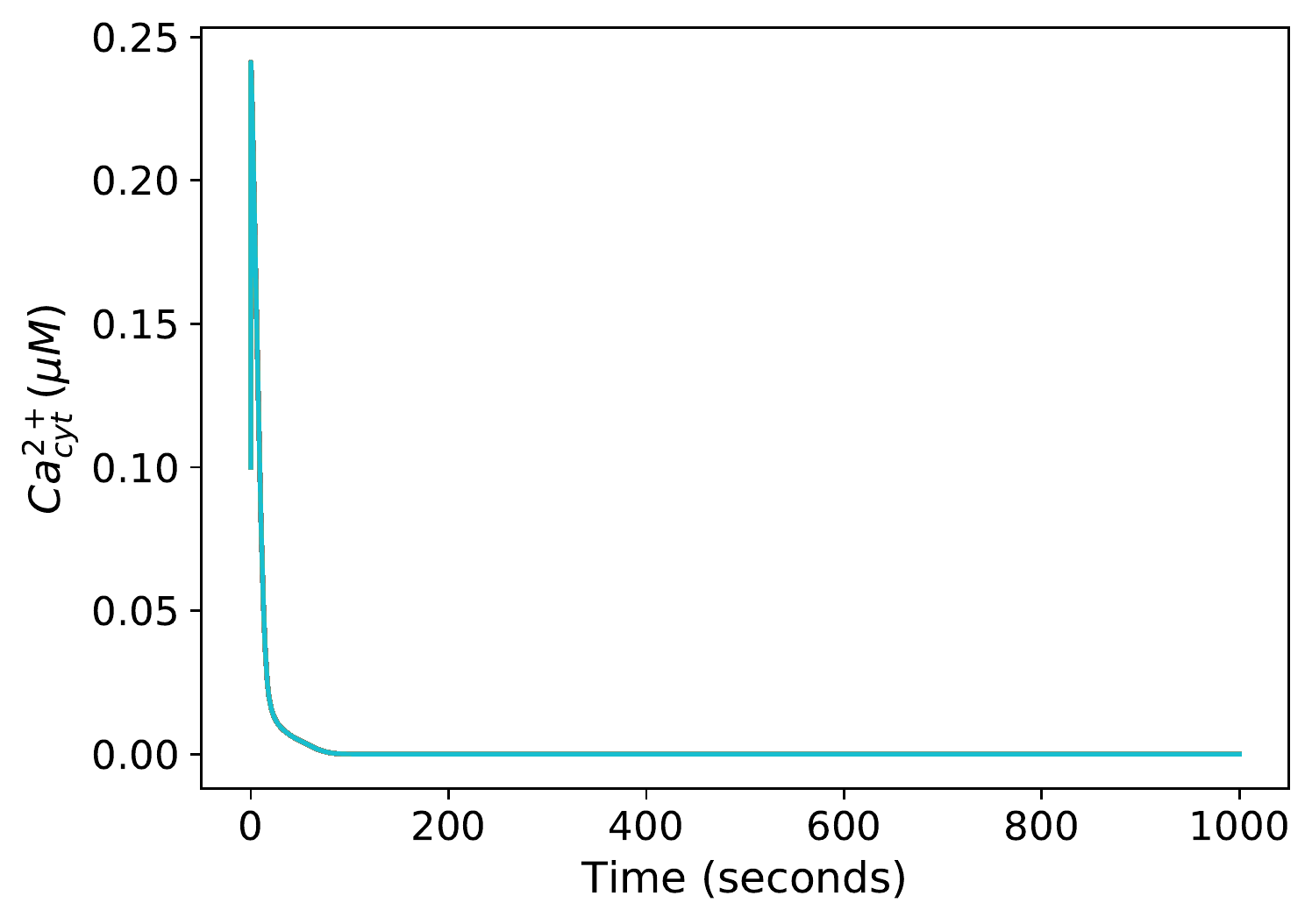}
  \caption{Case I: $Q$=0.0 and Stimulus=0.0}
  \label{F4a}
  \end{subfigure}%
\begin{subfigure}{0.44\textwidth}
  \centering
  \includegraphics[scale=0.45]{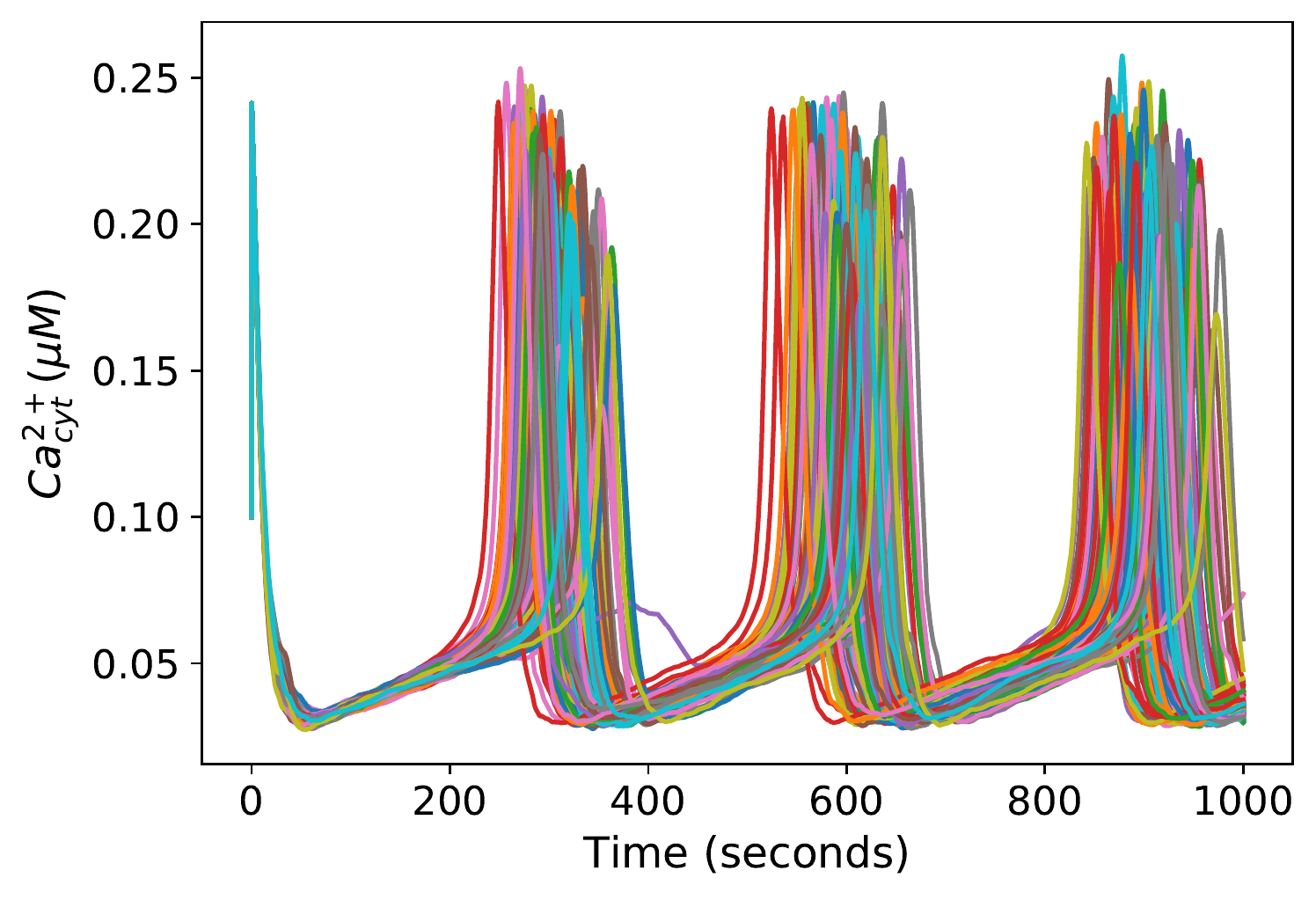}
  \caption{Case II: $Q$=0.0 and Stimulus is 0.17}
  \label{F4b}
  \end{subfigure}
\begin{subfigure}{0.44\textwidth}
  \centering
  \includegraphics[scale=0.45]{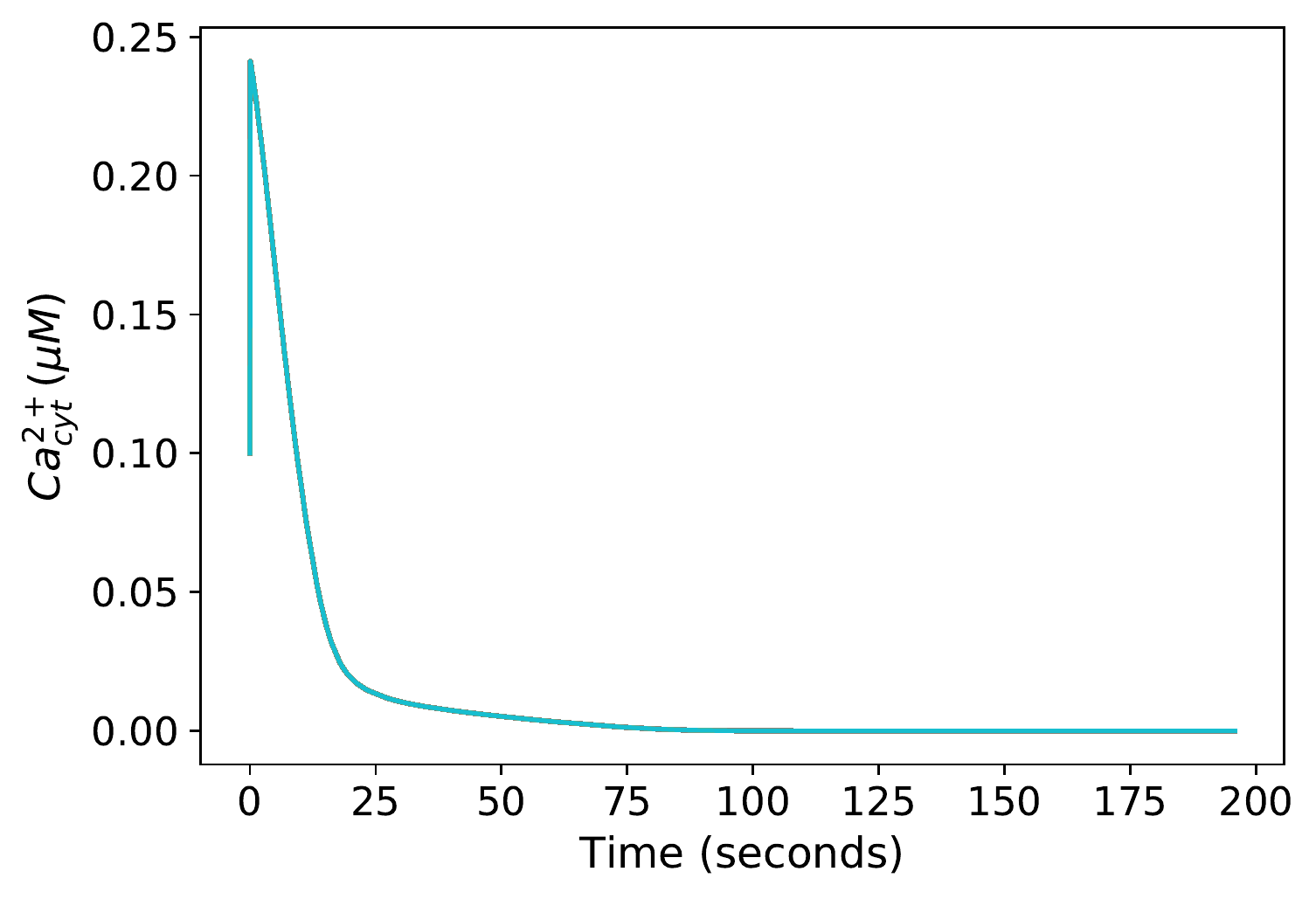}
  \caption{Case III: $Q$=0.18 and Stimulus=0.0}
  \label{F4c}
  \end{subfigure}%
\begin{subfigure}{0.44\textwidth}
  \centering
  \includegraphics[scale=0.45]{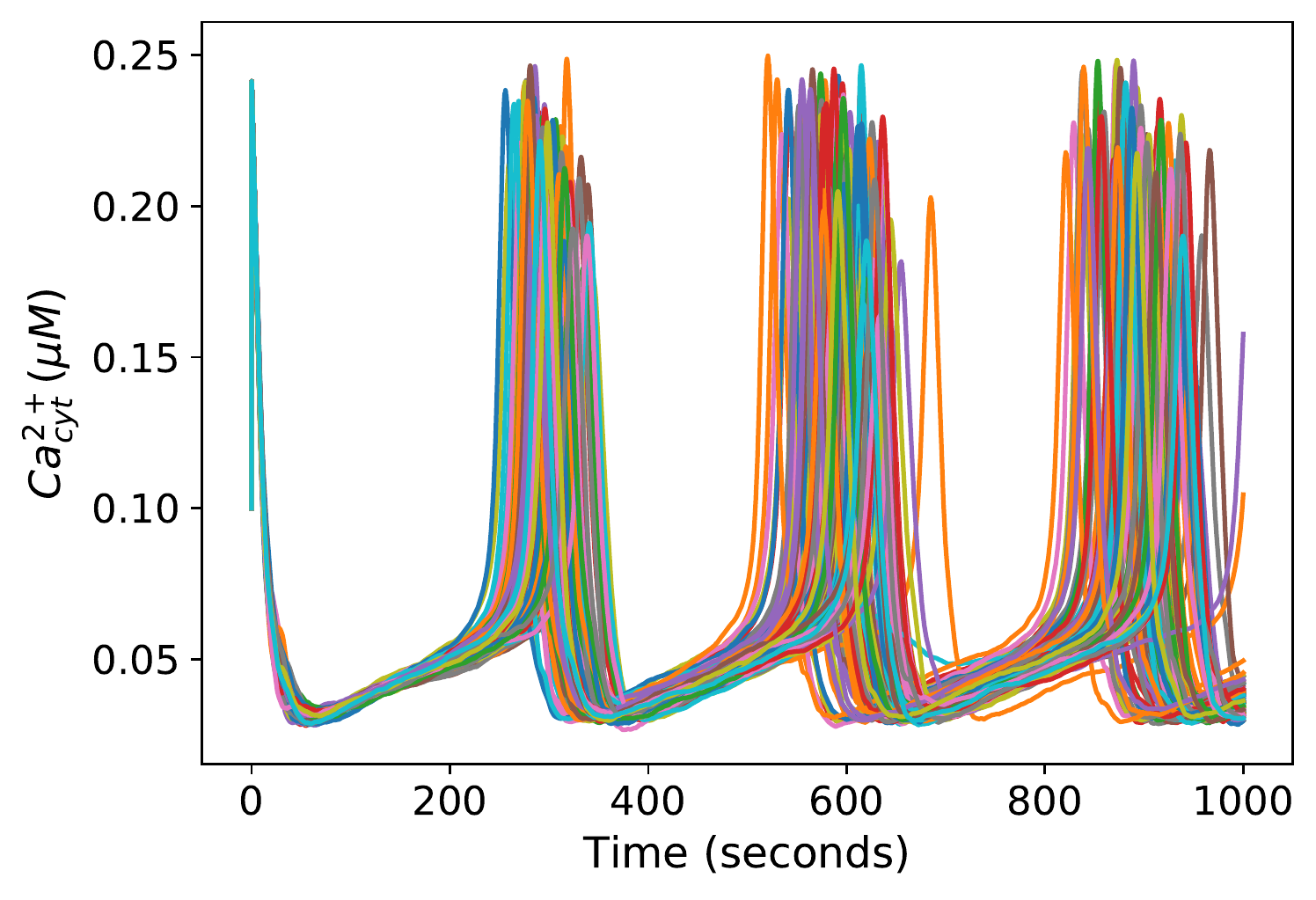}
  \caption{Case IV: $Q$=0.05 and Stimulus=0.17}
  \label{F4d}
  \end{subfigure}
\begin{subfigure}{0.44\textwidth}
  \centering
  \includegraphics[scale=0.45]{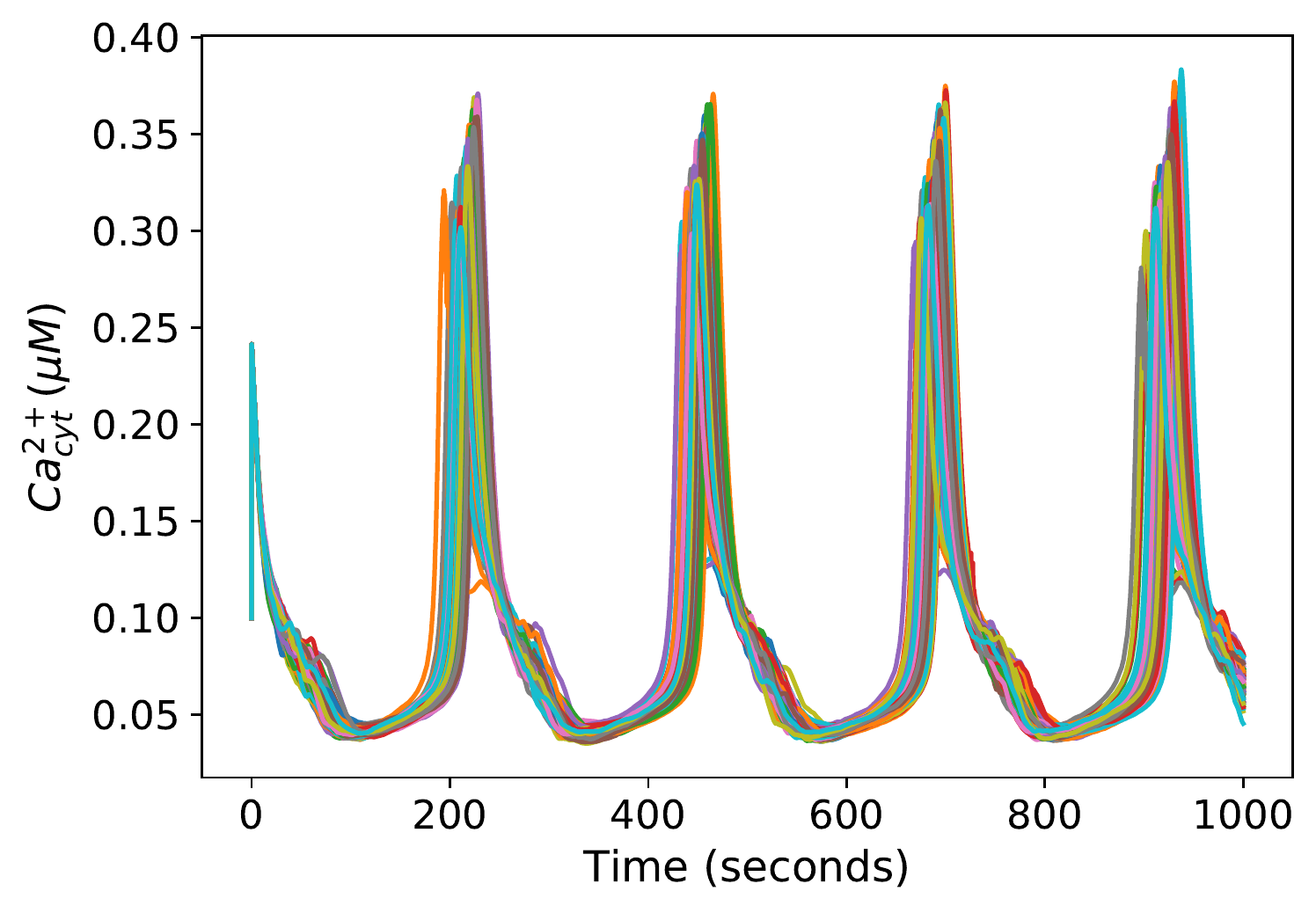}
  \caption{Case IV: $Q$=0.45 and Stimulus=0.17}
  \label{F4e}
  \end{subfigure}%
\begin{subfigure}{0.42\textwidth}
  \centering
  \includegraphics[scale=0.45]{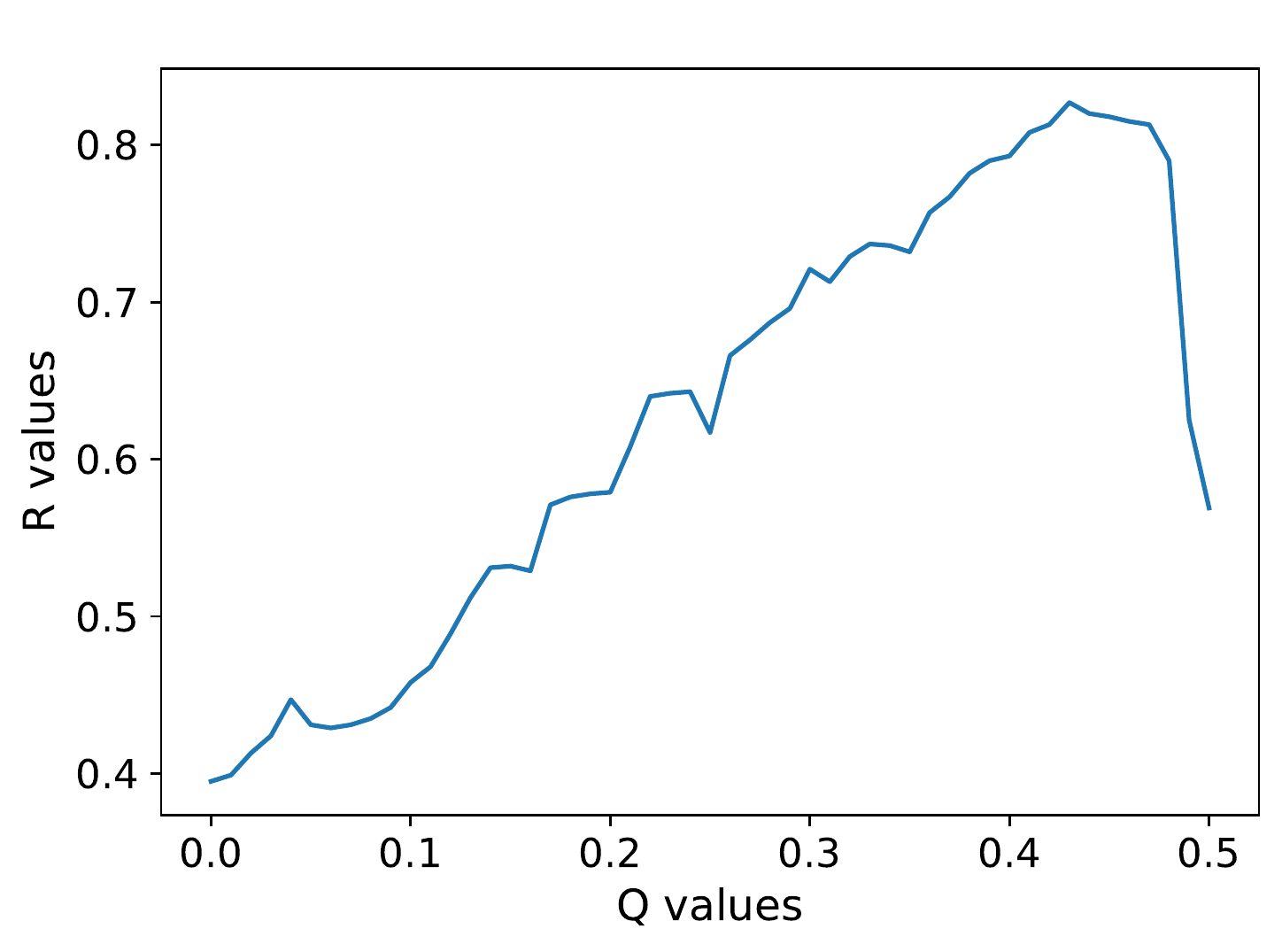}
  \caption{Variation of $R$ with $Q$ for Case IV}
  \label{F4f}
  \end{subfigure}%
\caption{Comparative simulations among four cases. (a) Case I: calcium dynamics in all the astrocytes are in a steady-state and the absence of oscillations is observed when both coupling strength and stimulus are kept zero. (b) Case II: stimulus of 0.17$\mu M sec^{-1}$ amplitude is given in the absence of coupling, calcium concentration starts fluctuating with varying amplitudes. (c) Case III: in the absence of stimulus and by varying the coupling strength, calcium dynamics achieve higher concentration once and then drop down for a longer period. Varying $Q$ does not make any change in cytosolic calcium dynamics. (d) and (e) belongs to Case IV: at a constant stimulus value of 0.17 $\mu M sec^{-1}$, coupling strength is varied from 0.05 to 0.50. As we move from $Q$=0.05 to 0.50, the level of synchronization is fluctuating, and cytosolic calcium shows the best synchrony result at $Q$=0.45. (f) $R$ dependence on $Q$: $R$ increases with the increase in coupling strength $Q$ till its value of 0.48 and starts decreasing thereafter.} 
\label{F4} 
\end{figure}

Next, we simulated a coupled A-N network of 100 units with random stimulus to neurons in following four scenarios: In case I, no stimulus is given and units are uncoupled (Q=0). In this case, no interesting dynamics is observed and all units quickly go into steady state as shown in Fig.~\ref{F4a}. In case II, units are kept uncoupled (Q=0) but stimulus is given to each cell with amplitude = 0.17 $\mu M sec^{-1}$. Here, the system oscillates as expected but with little synchrony as evident from Fig.~\ref{F4b}. Further, in case III, coupling strength $Q$ is fixed to 0.45 and no stimulus is given. It is visible that the system does not go into an oscillatory state, therefore, it can be deduced that glutamate may induce astrocytic calcium dynamics ~\citep{zur2006role}. The results are presented in Fig.~\ref{F4c}. In case IV, coupling strength (Q) is varied as shown in Fig.~\ref{F4d} and ~\ref{F4e} and stimulus is fixed to 0.17 $\mu M sec^{-1}$.  A transition to synchronization in intercellular calcium concentrations is observed when value of Q is shifted from 0.05 in Fig.~\ref{F4d} to 0.45 in Fig.~\ref{F4e}. 
Lastly, R is plotted with Q in Fig.~\ref{F4f} for fixed stimulus amplitude = 0.17 $\mu M sec^{-1}$. It is clearly evident that as Q increases, R also increases and reaches to a maximum value of around 0.82 at Q=0.45. Afterwhich it starts decreasing with further increase in Q. Thereby, suggesting the existence of optimized value of Q where the system is most synchronized. For further analysis, values of Q and stimulus corresponding to best case scenario of synchronization were chosen and fixed at 0.45 and 0.17 $\mu M sec^{-1}$, respectively, unless otherwise specified.  The above discussion is summarized as below.

\setlist{nolistsep}
\begin{itemize}
\item[] Case I: coupling strength, $Q$=0.0 and stimulus is 0.0
\item[] Case II: coupling strength, $Q$=0.0 and stimulus  is 0.17 
\item[] Case III: coupling strength, $Q$=varied and stimulus  is 0.0
\item[] Case IV: coupling strength, $Q$=varied and stimulus  is 0.17
\end{itemize}

\subsection{Effect of stimulus given to a fraction of A-N units on other units:}

\begin{figure}[H]
\begin{subfigure}{0.44\textwidth}
  \centering
  \includegraphics[width=0.92\linewidth]{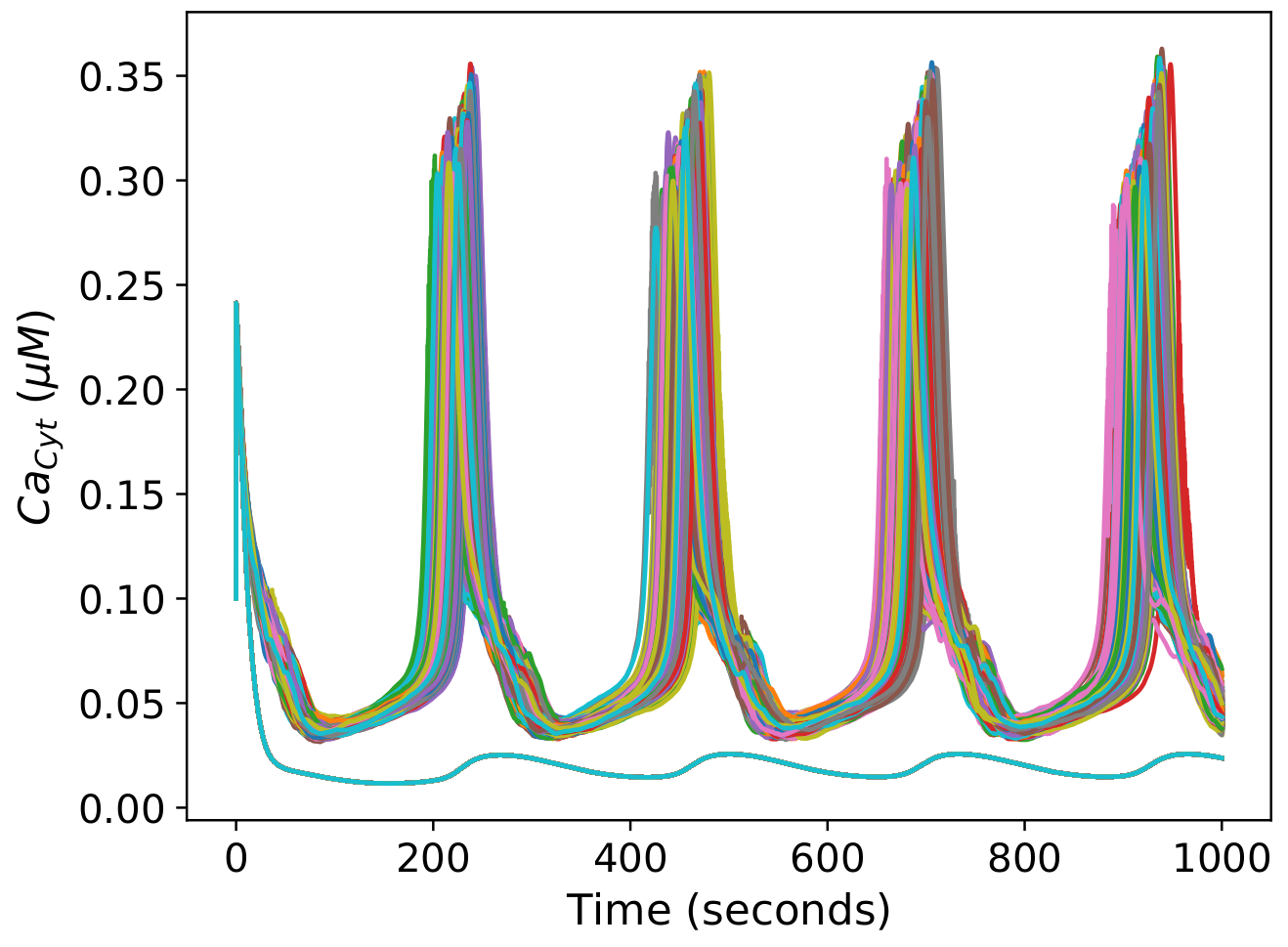}
  \caption{Q=0.45}
  \label{F5a}
  \end{subfigure}%
\begin{subfigure}{0.44\textwidth}
  \centering
  \includegraphics[width=0.92\linewidth]{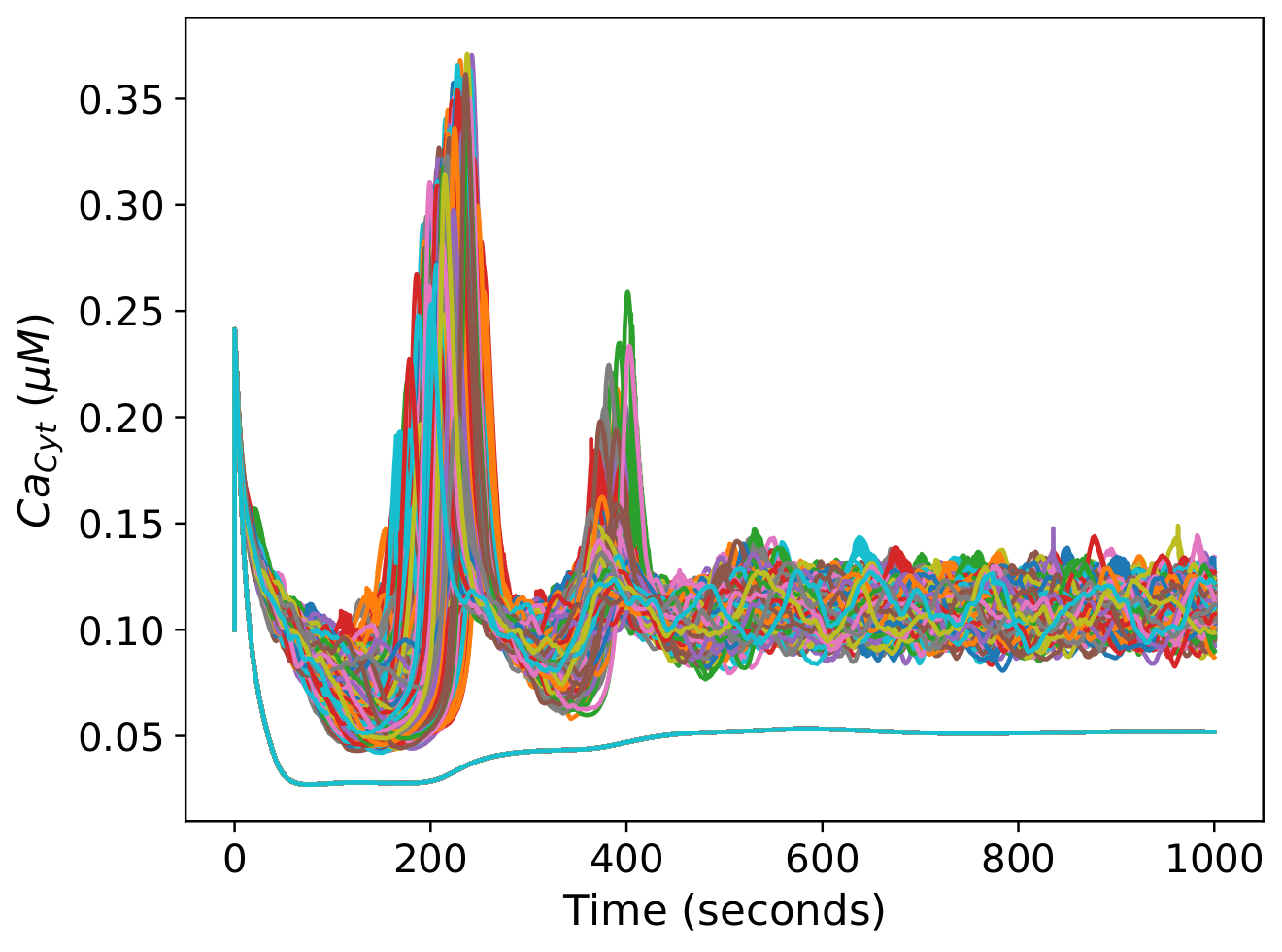}
  \caption{Q=0.55}
  \label{F5b}
  \end{subfigure}
\begin{subfigure}{0.44\textwidth}
  \centering
  \includegraphics[width=0.92\linewidth]{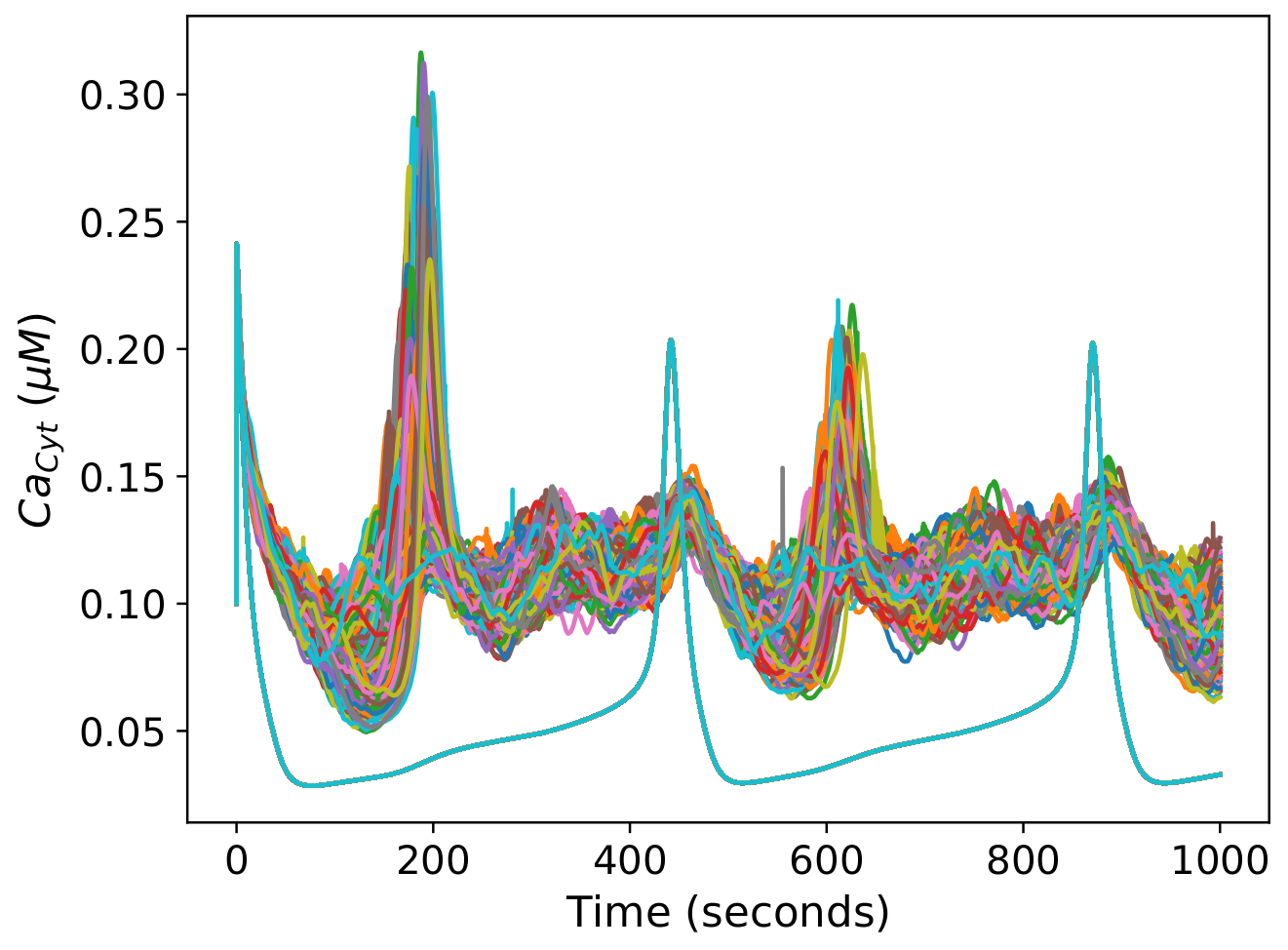}
  \caption{Q=0.56}
  \label{F5c}
  \end{subfigure}%
\begin{subfigure}{0.44\textwidth}
  \centering
  \includegraphics[width=0.92\linewidth]{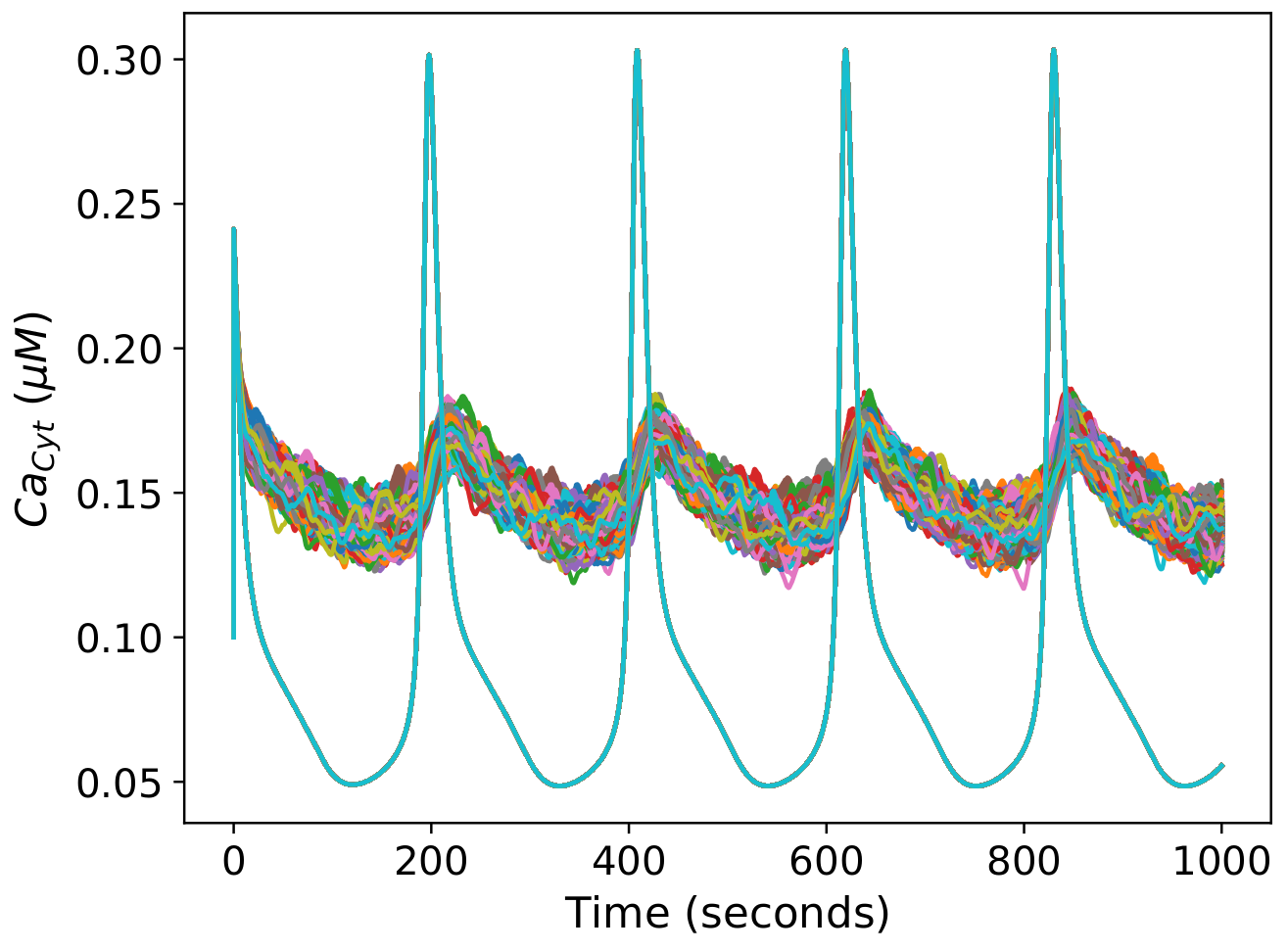}
  \caption{Q=0.64}
  \label{F5d}
  \end{subfigure}
\begin{subfigure}{0.44\textwidth}
  \centering
  \includegraphics[width=0.92\linewidth]{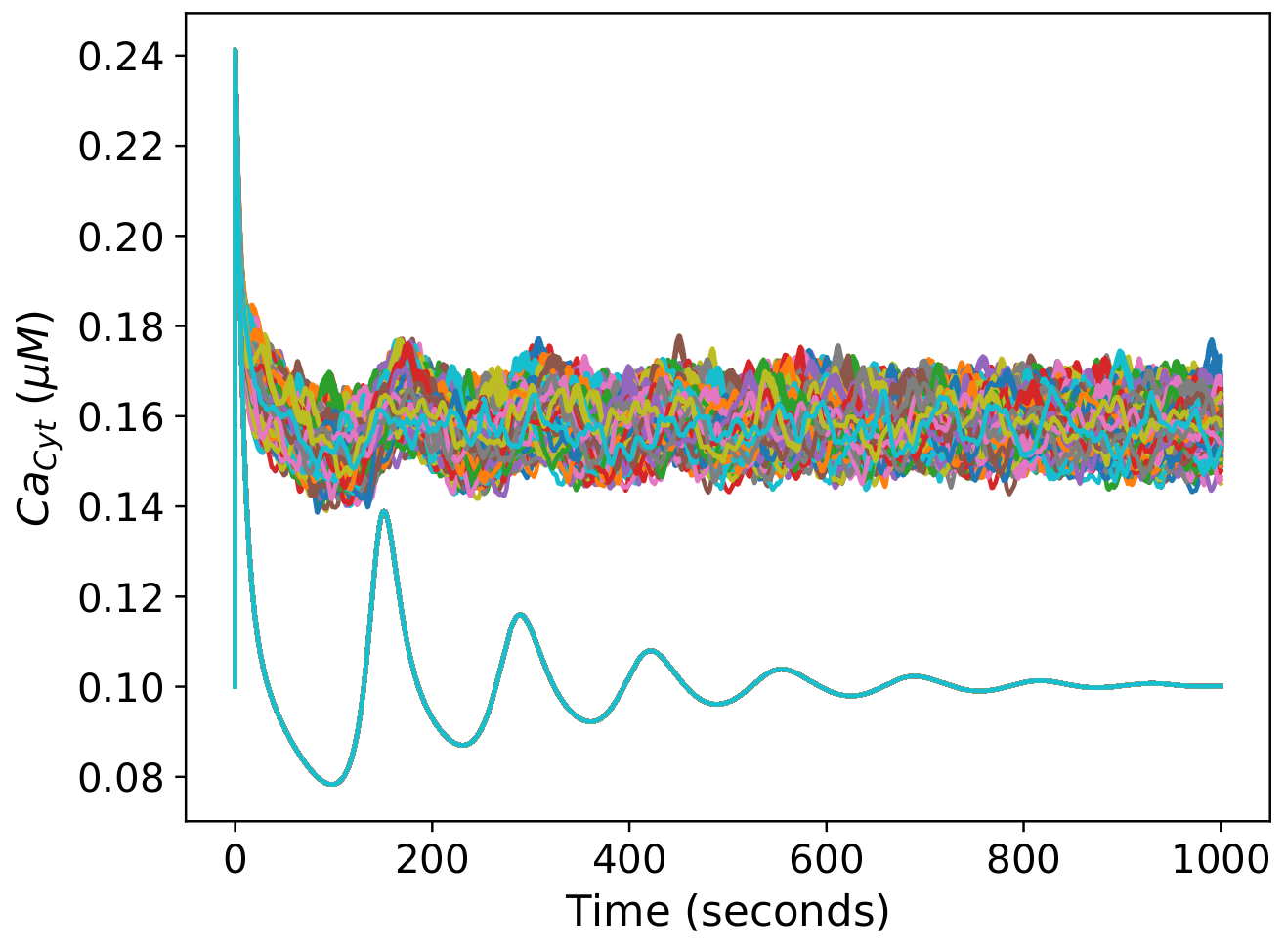}
  \caption{Q=0.65}
  \label{F5e}
  \end{subfigure}%
\begin{subfigure}{0.44\textwidth}
  \centering
  \includegraphics[width=0.92\linewidth]{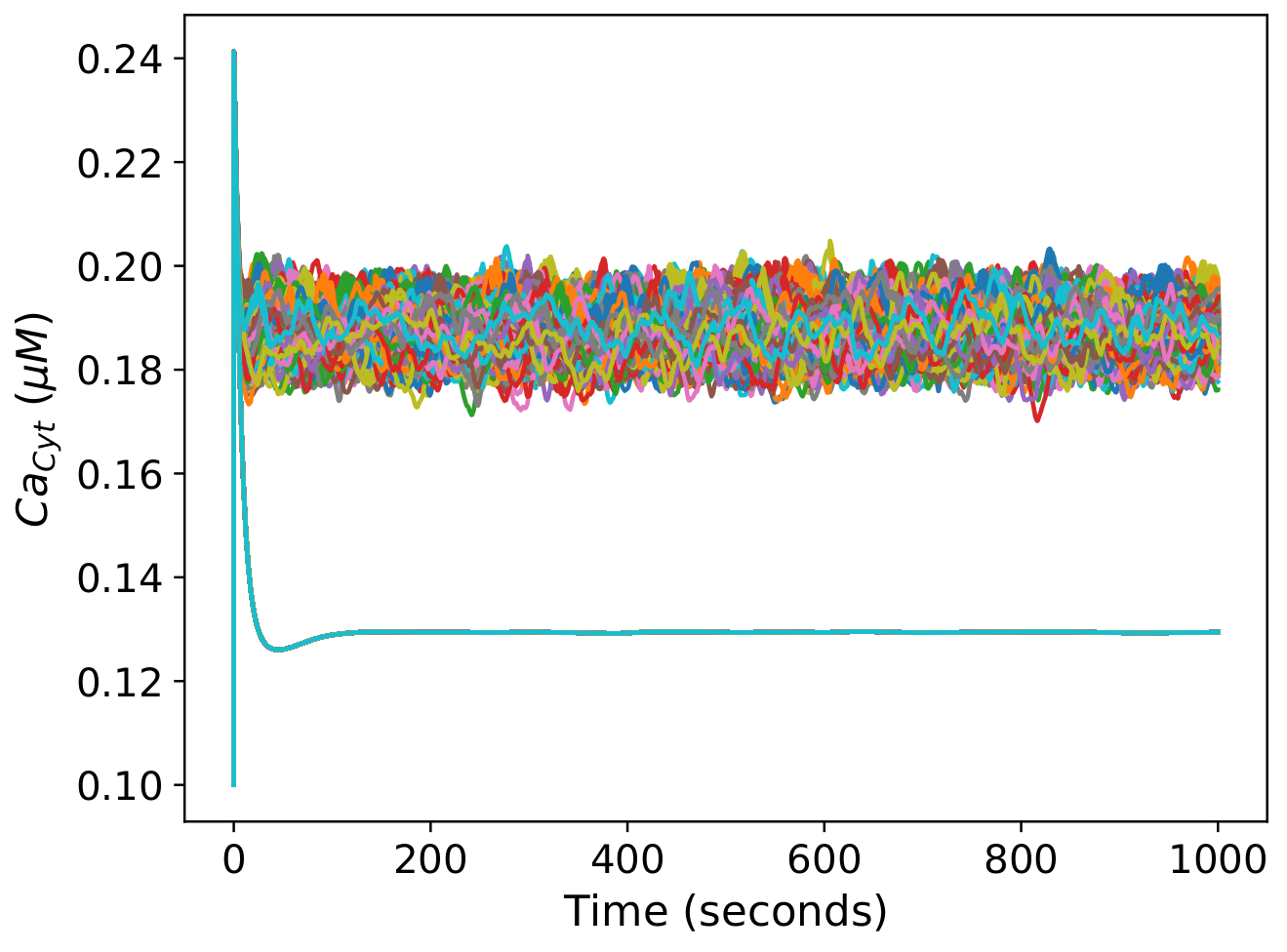}
  \caption{Q=0.70}
  \label{F5f}
  \end{subfigure}
\caption{(a), (b), (c) and (d) represent the oscillatory behaviour of $Ca^{2+}$ in astrocytes, where stimulus is given to 800 units out of 1,000 units, respectively. (a) oscillations of higher and lower amplitude can be seen with difference in SUs and NUs. Astrocytic calcium in SUs are oscillating with higher amplitude at Q=0.55. Whereas, oscillations of lower amplitude corresponds to NUs. (b) astrocytic $Ca^{2+}$ of both units converges to steady state with absence of oscillations at Q=0.55. (c) and (d) explains the specific range of coupling strength calcium of NUs also starts oscillating with much higher amplitude than others. (e) and (f) steady state position of astrocytic $Ca^{2+}$ of A-N units, when Q is increased to higher values (0.65 or more).}
\label{F5} 
\end{figure}

In this section, in an all-to-all connected network, fraction of units (eg. 800, in Fig ~\ref{F5a}) out of total units (1,000) were given stimulus (referred as, stimulated unit (SU)). The  effect of increasing coupling strength (Q) was observed on calcium dynamics in non-stimulated unit (NU). As evident from ~\ref{F5a}, for Q=0.45, system shows two different amplitudes of oscillations in SUs and NUs of similar frequency. The peaks in NUs lag behind the peaks in SUs. Despite the difference in amplitude, the units are fairly synchronized. NUs are shown in blue. As Q is increased, the synchronized behavior in both SUs and NUs starts to fade (~\ref{F5b}). On increasing Q further, large amplitude of oscillations among NUs start to emerge (~\ref{F5c}), although, the system is still not in a synchronized state. On further increase in Q, synchronized oscillations among SUs and NUs emerge.  At further higher values of Q, astrocytic calcium in all units transit to a steady state.


\subsubsection{Phase plot of ensemble size and coupling strength}

\begin{figure}[H]
\centering
\includegraphics[height=7.0cm, width=10.0cm]{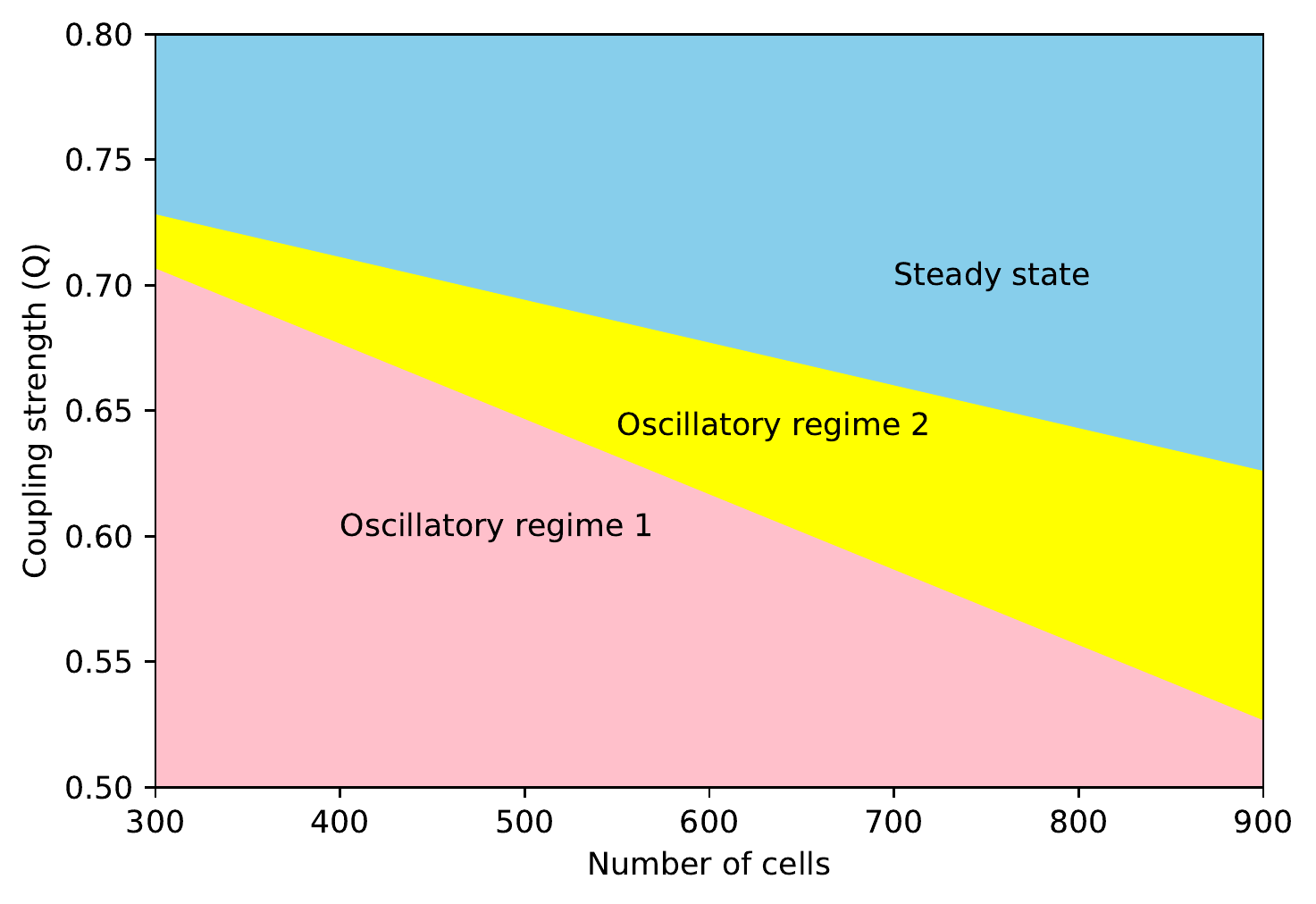}
\caption{Phase plot representing ranges of coupling strength, where the presence of oscillatory regimes were observed. Coupling strength is plotted on Y-axis with number of units getting stimulus on X-axis.}
\label{F6}
\end{figure}

Number of units receiving stimulus in A-N network was varied from 300 to 900 units in a difference of 50 as plotted in Fig.~\ref{F6}. Oscillatory regime 1 represent oscillations in NUs with very low amplitude. NUs shows sychronized oscillations in oscillatory regime 2. Steady state regime corresponds to that, all cells are in steady state.

It can be seen that emergent synchronized behaviour is observed, after at least 300 units starts receiving stimulus. The range at which this synchronizing behaviour is obserevd keeps on increasing with an increase in number of units getting stimulus. All the lower and higher ranges of coupling strength are shown in Table~\ref{Table2}. Here, we observed that there exists a critical value of number of units to be stimulated in order to achieve synchronization in NUs. Although, the exact reason for this criticality is not understood. 

As per equation (6) and (7), with increase in SUs, there will be a corresponding increase in $N_{out}$. Further, $N_{out}$ can diffuse freely and can enhance the release of glutamate, which leads to astrocytic calcium oscillations in non-stimulated A-N units. Therefore, the change in concentration of $N_{out}$ may be considered as an important factor for observed emergent dynamics.

\begin{table*}[h]
\begin{center}
\caption{Range of Q values giving rise to oscillatory behaviour in units not receiving stimulus.}
\label{Table2}
\begin{tabular}{ |p{5cm}|p{5cm}|p{5cm}|}
\hline
\textbf{Number of cells getting stimulus(*)} & \textbf{Q1 (lower value)} & \textbf{Q2 (higher value)} \\
\hline
300 & 0.72 & 0.73 \\
\hline
350 & 0.70 & 0.72 \\
\hline
400 & 0.68 & 0.71 \\
\hline
450 & 0.66 & 0.70  \\
\hline
500 & 0.64 & 0.69 \\
\hline
550 & 0.63 & 0.68 \\
\hline
600 & 0.60 & 0.68 \\
\hline
650 & 0.60 & 0.67 \\
\hline
700 & 0.58 & 0.66 \\
\hline
750 & 0.57 & 0.65 \\
\hline
800 & 0.56 & 0.64 \\
\hline
850 & 0.55 & 0.63 \\
\hline
900 & 0.54 & 0.63  \\
\hline
\end{tabular}
\end{center}
\end{table*}

\section{Discussion}

Astrocytes are now being increasingly recognized as an important component in information processing ~\citep{araque2008astrocytes}. Their role in synaptic modulation is well known ~\citep{perea2009synaptic}, but their function at network level in achieving the same is still relatively less understood. To study the same, we  developed the concept of A-N unit and studied  an ensemble of these units in the form of a coupled newtork of these is to understand and check the role of NO in coupled astrocytes. How does it lead to a meaningful connection between ensemble of astrocytes? This work is designed on the basis of a single astrocytic model given by Lavrentovich \& Hemkin ~\citep{lavrentovich2008mathematical}. While their model was limited to only a single astrocytic cell, the current model goes a step further by connecting astrocytes with neurons and developing ensemble of astrocyte-neuronal units $via$ NO diffusion molecule. The existence and importance of NO in astrocytes have been verified in various experimental papers ~\citep{buskila2007enhanced, contestabile2012neuronal}. Therefore, it becomes an interesting idea to explore the potential roles of NO in coupled A-N units. NO interacts with glutamate release in a complex manner. Not only the presence of nitric oxide can increase glutamate production ~\citep{wang2017nitric}, it can also stimulate vesicle release process without depending upon calcium signalling ~\citep{meffert1996nitric}. Thus, it can make a neuron take part in information processing by bypassing the calcium dependent vessicle release mechanism in a network. As this process does not require firing of action potentials and other calcium dynamics, it can reduce the overall energy demand of the neuron ~\citep{attwell2001energy, yu2018evaluating} .

Astrocytes on the other hand are known to consume about 15 to 30\% of total energy available to CNS ~\citep{wang2009astrocytic}. Energy consumed as number of ATPs/second for an excitatory neuron during action potential generation and propagation is around $2.65*10^{8}$ at 1 Hz ~\citep{yu2018evaluating}. For astrocytes during calcium dynamics, it is around $0.92*10^{8}$. In this study, we found that, NO facilitates synaptic transmission by inducing synchronized astrocytic calcium waves without the need for calcium dependent vesicle release mechanism in neurons not receiving stimuli. Based on the above arguments, the model predicts that intercellular synchronized calcium oscillations in astrocytes can aid in synaptic transmission in an energy efficient manner over long distances. 

Results of this work may also be examined, at least qualitatively, in the light of the ``glissandi" phenomenon, first reported in a culture study ~\citep{kuga2011large}, reporting that in a large network of hundreds of astrocytes, calcium oscillations in almost all the astrocytes become synchronized and propagate as a wave. The phenomenon was correlated with around 42\% decrease in infraslow fluctuations. Although the exact origins of infraslow potential are yet not clear, multiple factors for the same have been suggested. These include $K^{+}$ ion concentration, $Na^{+}$/$K^{+}$ pumps, ~\citep{krishnan2018origin}, coupling strength, structural connectivity ~\citep{ghosh2008noise} and ATP ~\citep{lHorincz2009atp}. 

Therefore, it can be safely assumed that a decrease in infraslow potential might be corelated with the decrease in neuronal energy consumption, and as argued before, the corresponding increase in energy consumption by astrocytic calcium dynamics is still less than the energy consumed by the neurons. The predicted results are worthy for further experimental analyses. In future studies, the detailed dynamics of energy consumption by neurons and astrocytes may be incorporated in this model, and the model may further be extended into a network of tripartite synapse to study the effect of NO on long term potentiation (LTP) and long term depression (LTD).

\begin{table*}[h]
\begin{center}
\caption{Values of parameters used in the model}
\label{Table1}
\begin{tabular}{ |p{3cm}|p{3cm}|p{3cm}|p{3cm}|p{3cm}|}
\hline
\textbf{Parameter} & \textbf{Value} & \textbf{Parameter} & \textbf{Value}\\
\hline
$k_{g}$    & 0.17 $s^{-1}$     & $\upsilon_{M3}$ & 40.0 $s^{-1}$      \\
\hline
$k_{out}$  & 0.5 $s^{-1}$      & $\upsilon_{M2}$ & 15.0 $\mu M s^{-1}$ \\
\hline
$k_{f}$    & 0.5 $s^{-1}$      & $k_{CaA}$       & 0.15 $\mu M$       \\
\hline
$v_{p}$    & 0.05 $\mu Ms^{-1}$ & $k_{CaI}$      & 0.15 $\mu M$    \\
\hline
$k_{p}$    & 0.47 $\mu M$      & $k_{ip3}$       & 0.1 $\mu M$      \\
\hline
$k_{deg}$  & 0.04 $s^{-1}$     & $k_{2}$         & 0.17 $\mu M$      \\
\hline
$k_{3}$    & 1.8 $s^{-1}$      & $\zeta$         & 0.4 $s^{-1}$       \\
\hline
$k_{pp}$   & 2.7 $s^{-1}$      & $g$             & 0.52 $\mu Ms^{-1}$  \\
\hline  
$k_{glut}$ & 0.2 $s^{-1}$      & n               & 2.02               \\
\hline
$\eta$     & 0.95 $s^{-1}$     & m               & 2.2               \\
\hline

\end{tabular}
\end{center}
\end{table*}

\bibliography{astrocyte3}
\bibliographystyle{unsrt}

\end{document}